\documentclass[twocolumn,prd,a4paper,final,nofootinbib,longbibliography]{revtex4-1}

\usepackage[dvipdfmx]{graphicx}
\usepackage[dvipdfmx]{hyperref}
\usepackage{amsmath,amssymb}
\usepackage{bm} 
\usepackage{mathtools}

\usepackage{color}
\usepackage{ulem}

\begin{document}

\title{Numerical study of the SWKB condition of novel classes of exactly solvable systems}
\author{Yuta Nasuda}
\email{y.nasuda.phys@gmail.com}
\affiliation{Department of Physics, Graduate School of Science and Technology, Tokyo University of Science, Noda, Chiba 278-8510, Japan}

\author{Nobuyuki Sawado}
\email{sawadoph@rs.tus.ac.jp}
\affiliation{Department of Physics, Graduate School of Science and Technology, Tokyo University of Science, Noda, Chiba 278-8510, Japan}

\date{\today}

\begin{abstract}

The supersymmetric WKB (SWKB) condition is supposed to be exact for all known exactly solvable quantum mechanical systems with the shape invariance. 
Recently, it was claimed that the SWKB condition was not exact for the extended radial oscillator, whose eigenfunctions consisted of the the exceptional orthogonal polynomial, even the system possesses the shape invariance.
In this paper, we examine the SWKB condition for the two novel classes of exactly solvable systems:  one has the multi-indexed Laguerre and Jacobi polynomials as the main parts of the eigenfunctions, and the other has the Krein--Adler Hermite, Laguerre and Jacobi polynomials.
For all of them, one can always remove the $\hbar$-dependency from the condition, and it is satisfied with a certain degree of accuracy. 
\end{abstract}


\maketitle

\section{Introduction}
\label{sec:introduction}
The well-known WKB quantization condition is given by
\begin{equation}
\int_{x_{\mathrm{L}}}^{x_{\mathrm{R}}} \sqrt{\mathcal{E}_n - V(x)}~\mathrm{d}x 
= \left( n + \frac{1}{2} \right) \pi\hbar
\qquad (n \in \mathbb{Z}_{\geqslant 0}) ~,
\label{eq:WKB}
\end{equation}
in which $x_{\mathrm{L}}$ and $x_{\mathrm{R}}$ are the ``turning points''; $V(x_{\mathrm{L}}) = V(x_{\mathrm{R}}) = \mathcal{E}_n$.
In the context of supersymmetric quantum mechanics (SUSY QM)~\cite{Witten:1981nf,witten1982constraints,Cooper:1994eh}, 
the potential $V(x)$ is formally given by the groundstate eigenfunction of the system $\phi_0(x)$ as 
\begin{equation}
V(x) = \hbar^2\left[ ( \partial_x\ln|\phi_0(x)| )^2 + \partial_x^2\ln|\phi_0(x)| \right] ~.
\end{equation}
It should be emphasized that the above potential $V(x)$ corresponds to the \textit{vanishing groundstate energy}, $\mathcal{E}_0=0$.

On the other hand, a WKB-like condition in SUSY QM (the SWKB condition) proposed by Comtet \textit{et al}.~\cite{Comtet:1985rb} reads 
\begin{equation}
\int_a^b \sqrt{\mathcal{E}_n - \left( \hbar\partial_x\ln|\phi_0(x)| \right)^2}~\mathrm{d} x = n\pi\hbar \qquad (n \in \mathbb{Z}_{\geqslant 0}) ~,
\label{eq:SWKB}
\end{equation}
where $a,b$ are the roots of $\left( \hbar\partial_x\ln|\phi_0(x)| \right)^2 = \mathcal{E}_n$.
This condition is exact for the groundstate by construction.
For all conventional shape-invariant potentials~\cite{gendenshtein1983derivation}, 
it has been demonstrated that the SWKB condition \eqref{eq:SWKB} reproduces the exact bound-state spectra by 
Dutt et al.~\cite{Dutt:1986pi}. 
It is well known that the shape invariance is a sufficient (but not a necessary) condition for the exact solvability of the Schr\"{o}dinger equation.  
A natural question that arises here is whether or not the shape invariance of the potential is necessary in order that the SWKB condition reproduces the exact bound-state spectrum. 
This was already discussed by Khare et al.~\cite{Khare:1989zy} with the Ginocchio potential and also a potential 
which is isospectral to the 1D harmonic oscillator, both of which are exactly solvable but are not shape invariant. 
They arrived at the conjecture that the shape invariance could be a necessary condition for the exactness of the SWKB condition.  

As was mentioned in Ref.~\cite{Comtet:1985rb}, any qualitative argument or proof of the exactness of the SWKB condition is still absent at present, and then the conjecture of Ref.~\cite{Khare:1989zy} has not been proved yet. 
In order to come closer to the answer of this question, it is worthwhile to examine the SWKB condition for new examples which are all exactly solvable but are not shape-invariant.
Khare and Varshni \cite{Khare:1989zy} reported the non-exactness of the SWKB condition for the above-mentioned potentials, which are not shape invariant. 
Around the same time, it was also pointed out that the SWKB condition is neither exact nor never worse than 
WKB for the Abraham--Moses systems~\cite{abraham1980changes}, which are solvable, by DeLaney \textit{et al}.~\cite{DeLaney:1990uj}. 

Recently, Bougie \textit{et al}.~\cite{Bougie:2018lvd} showed that the SWKB condition is not always exact for shape-invariant potentials. 
Their claim that the shape invariance is not sufficient condition for the SWKB exactness is suggestive.
However, their analysis is wrong because it lacks the proper treatment of $\hbar$. 
We now realize that there still is a possibility that the shape invariance does not guarantee the exactness of the SWKB condition. 
Section \ref{sec:MI} is devoted to a further numerical analysis on the novel class of shape-invariant potentials.

In Refs.~\cite{adhikari1988higher,dutt1991supersymmetry}, the authors tried to understand the non-exactness of the SWKB condition for the systems without shape invariance in a context of the semi-classical regime, where the SWKB condition is obtained from the standard WKB formula.
Although most of the literature employs the unit of $\hbar=1$, to simplify the analyses, we retain $\hbar$ in this paper for rigorous discussions.
In fact, as we shall show, $\hbar$ is removed completely from the SWKB condition~\eqref{eq:SWKB}.
It apparently means that the SWKB exactness should not be discussed in the context of the semi-classical regime of the quantum system.

Here, a fundamental question arises: what does the SWKB condition imply?  
The exactness of the condition might indicate that the potential belongs to the class of conventional shape-invariant ones.
Breaking of the SWKB condition somehow reflects the discrepancy between the potential 
of our concern and the corresponding shape-invariant system, which might be depicted by an \textit{unknown factor}.
For the moment, we have only a vague notion to realize it. 
Thus, it is worth performing the numerical study of several novel classes of solvable systems to get an insight for the factor. 

In this paper, we study the systems with the multi-indexed Laguerre and Jacobi polynomials~\cite{Odake:2011jj} as the main parts of the eigenfunctions and the Krein--Adler systems~\cite{Kre57,adler1994modification}. 
They are obtained by deforming the harmonic oscillator (H), the radial oscillator, or the pseudo-harmonic oscillator, (L) and the P\"{o}schl--Teller potential (J) (we set $2m=1$): 
\begin{align}
&\mathcal{H}^{(\ast)} = -\hbar^2\partial_x^2 + V^{(\ast)}(x) ~,
\label{eq:ES_ham} \\
&V^{(\ast)}(x) = \left\{
	\begin{array}{l}
	\omega^2x^2 - \hbar\omega \hfill \mathrm{\ast=H} \\
	\omega^2x^2 + \dfrac{\hbar^2g(g-1)}{x^2} - \hbar\omega(2g+1) \\
	\hspace{5cm} \mathrm{\ast=L} \\
	\dfrac{\hbar^2 g(g-1)}{\sin^2x} + \dfrac{\hbar^2 h(h-1)}{\cos^2x} - \hbar^2(g+h)^2 \\
	\hspace{5cm} \mathrm{\ast=J}
	\end{array}
\right. 
\nonumber
\end{align}
with the Schr\"{o}dinger equations
\begin{align}
&\mathcal{H}^{(\ast)}\phi_n^{(\ast)}(x) = \mathcal{E}_n^{(\ast)}\phi_n^{(\ast)}(x) ~\quad (n\in\mathbb{Z}_{\ge0}),
\label{eq:ES_Sch} \\
&\mathcal{E}_n^{(\ast)} = \left\{
        \begin{array}{lcl}
        2n\hbar\omega &\qquad& \mathrm{\ast=H} \\
        4n\hbar\omega && \mathrm{\ast=L} \\
        4\hbar^2 n(n+g+h) && \mathrm{\ast=J} 
        \end{array}
\right. ~, \nonumber\\
&\phi_0^{(\ast)} (x) = \left\{
        \begin{array}{lcl}
        \mathrm{e}^{-\frac{\omega x^2}{2\hbar}}, &\qquad& \mathrm{\ast=H} \\
        \mathrm{e}^{-\frac{\omega x^2}{2\hbar}}x^g, && \mathrm{\ast=L} \\
        (\sin x)^g(\cos x)^h, && \mathrm{\ast=J} 
        \end{array}
\right. ~. \nonumber
\end{align}
The shape invariance is achieved by $g\to g+1$ and $h\to h+1$.
We note that the significant parts of the eigenfunctions $\{ \phi_n^{(\ast)}(x) \}$ are described by Hermite polynomials 
$H_n$ for the harmonic oscillator (H), Laguerre polynomials $L_n^{(\alpha)}$ for the radial oscillator (L) 
and Jacobi polynomials $P_n^{(\alpha,\beta)}$ for the P\"{o}schl--Teller potential (J), respectively. 
We shall see that for both of the multi-indexed and the Krein--Adler systems, the SWKB condition is exact only for the groundstates $n=0$, 
which is obvious by construction, but they could be a good approximation especially for sufficiently higher excited states.
We shall also discuss the mathematical implications of the results in Sec.~\ref{sec:discussion}.

\section{Multi-indexed systems and the SWKB condition}
\label{sec:MI}

In this section, we examine the SWKB condition for the novel class of shape-invariant potentials, i.e., the multi-indexed systems. 

Bougie \textit{et al}.~\cite{Bougie:2018lvd} discuss non-exactness of the SWKB condition for the additive shape-invariant potentials. 
The authors employed the extended radial oscillator, which was equivalent to the exceptional Laguerre or the type II $X_1$-Laguerre polynomial~\cite{gomez2004supersymmetry,gomez2009extended,gomez2010extension,Quesne_2008,Odake:2009zv,Odake:2010zz,Sasaki_2010}, to show that the SWKB condition was not exact for all additive shape-invariant potentials. 
The extended radial oscillator can be seen as a special case of the multi-indexed systems.
We note that they alleged that the additive shape invariance was realized for the parameters $a_i$ such that 
$a_{i+1}=a_i+\hbar$ (also in Refs.\cite{bougie2010generation,bougie2012supersymmetric,bougie2015generation,mallow2020inter}) 
and expanded the superpotential $W$ in power of $\hbar$ where the conventional shape-invariant superpotentials are 
the lowest order $\mathcal{O}(\hbar^0)$. Note stress that the SWKB condition has no dependency on $\hbar$ and no room for
their expansion parameter $\hbar$. The analysis is thus quite questionable. (Another drawback of their analysis is discussed in Appendix A.)

We carry out a further numerical analysis not only on the extended radial oscillator but also on the multi-indexed systems, which are more general cases of the novel shape-invariant potentials.

\subsection{Multi-indexed Laguerre/Jacobi systems}
The multi-indexed Laguerre and Jacobi polynomials~\cite{Odake:2011jj} are obtained through deformations of two of the three exactly solvable systems \eqref{eq:ES_ham}
via the virtual-state wavefunctions $\{ \varphi_n(x) \}$, which are defined as
\begin{align}
\varphi_n^{\mathrm{(L),I}}(x) &\coloneqq 
\mathrm{e}^{\frac{z}{2}} z^{\frac{g}{2}} L_n^{(g-\frac{1}{2})}(-z) ~, \\
\varphi_n^{\mathrm{(L),II}}(x) &\coloneqq 
\mathrm{e}^{-\frac{z}{2}} z^{\frac{1-g}{2}} L_n^{(\frac{1}{2}-g)}(z) ~, \\
\varphi_n^{\mathrm{(J),I}}(x) &\coloneqq 
\left( \dfrac{1-y}{2} \right)^{\frac{g}{2}} \left( \dfrac{1+y}{2} \right)^{\frac{1-h}{2}} P_n^{(g-\frac{1}{2},\frac{1}{2}-h)}(y) ~, \\
\varphi_n^{\mathrm{(J),II}}(x) &\coloneqq 
\left( \dfrac{1-y}{2} \right)^{\frac{1-g}{2}} \left( \dfrac{1+y}{2} \right)^{\frac{h}{2}} P_n^{(\frac{1}{2}-g,h-\frac{1}{2})}(y) ~,
\end{align}
where
\begin{equation}
\xi \equiv \sqrt{\frac{\omega}{\hbar}}x ~,~~~
z \equiv \xi^2 ~,~~~
y \equiv \cos 2x ~,
\end{equation}
and the parameters $g,h$ must satisfy
\begin{align}
\text{L:}\qquad g &> \mathrm{max}\left\{ N+\frac{3}{2}, d_j^{\mathrm{II}}+\frac{1}{2} \right\} ~, \\
\text{J:}\qquad g &> \mathrm{max}\left\{ N+2, d_j^{\mathrm{II}}+\frac{1}{2} \right\} ~,\nonumber \\
h &> \mathrm{max}\left\{ M+2, d_j^{\mathrm{I}}+\frac{1}{2} \right\} ~.
\end{align}
We employ the virtual-state wavefunctions with $n\in \mathcal{D}=\mathcal{D}^{\mathrm{I}}\cup\mathcal{D}^{\mathrm{II}}=\{ d_1^{\mathrm{I}},\ldots,d_M^{\mathrm{I}} \}\cup\{ d_1^{\mathrm{II}},\ldots,d_N^{\mathrm{II}} \}$ and $d_1^{\mathrm{I}}<\cdots<d_M^{\mathrm{I}},d_1^{\mathrm{II}}<\cdots<d_N^{\mathrm{II}}\in\mathbb{Z}_{>0}$ as the seed solutions, and deform $\{\phi_n(x)\}$ through the multiple Darboux transformation to construct the multi-indexed systems.
The resulting deformed systems are
\begin{align}
&\mathcal{H}_{\mathcal{D}}^{\mathrm{(M,\ast)}} \coloneqq \mathcal{H}^{(\ast)} \nonumber \\
& - 2\hbar^2\partial_x^2\ln\left|\mathrm{W}\left[\varphi_{d_1^{\mathrm{I}}}^{\mathrm{(\ast),I}},\ldots,\varphi_{d_M^{\mathrm{I}}}^{\mathrm{(\ast),I}},\varphi_{d_1^{\mathrm{II}}}^{\mathrm{(\ast),II}},\ldots,\varphi_{d_N^{\mathrm{II}}}^{\mathrm{(\ast),II}}\right](x)\right|
\end{align}
with
\begin{align}
&\mathcal{H}_{\mathcal{D}}^{\mathrm{(M,\ast)}}\phi_{\mathcal{D};n}^{\mathrm{(M,\ast)}}(x) 
= \mathcal{E}_{\mathcal{D};n}^{\mathrm{(M,\ast)}}\phi_{\mathcal{D};n}^{\mathrm{(M,\ast)}}(x) 
\qquad (n \in \mathbb{Z}_{\geqslant 0}) ~, \\
&\phi_{\mathcal{D};n}^{\mathrm{(M,\ast)}}(x) 
\nonumber \\
&~~= \frac{\mathrm{W}\left[\varphi_{d_1^{\mathrm{I}}}^{\mathrm{(\ast),I}},
\ldots,\varphi_{d_M^{\mathrm{I}}}^{\mathrm{(\ast),I}},\varphi_{d_1^{\mathrm{II}}}^{\mathrm{(\ast),II}},
\ldots,\varphi_{d_N^{\mathrm{II}}}^{\mathrm{(\ast),II}},\phi_n^{(\ast)}\right](x)}{\mathrm{W}\left[\varphi_{d_1^{\mathrm{I}}}^{\mathrm{(\ast),I}},
\ldots,\varphi_{d_M^{\mathrm{I}}}^{\mathrm{(\ast),I}},\varphi_{d_1^{\mathrm{II}}}^{\mathrm{(\ast),II}},\ldots,\varphi_{d_N^{\mathrm{II}}}^{\mathrm{(\ast),II}}
\right](x)} ~,\\
&\mathcal{E}_{\mathcal{D};n}^{\mathrm{(M,\ast)}} = \mathcal{E}_n^{(\ast)} ~.
\label{eq:MIene}
\end{align}
in which $\ast=\mathrm{L,J}$ and 
\[
\mathrm{W}[f_1,\cdots,f_n](x)=\mathrm{det}\left( \partial_x^{j-1}f_k(x) \right)_{1\leqslant j,k\leqslant n}
\]
is the Wronskian. 
Especially, the groundstates are written as
\begin{align}
&\phi_{\mathcal{D};0}^{(\mathrm{M,L})}(x) = z^{\frac{M+N}{2}}
\nonumber \\ 
&\quad \times\frac{\mathrm{W}\left[\varphi_{d_1^{\mathrm{I}}}^{\mathrm{(L),I}},\ldots,\varphi_{d_M^{\mathrm{I}}}^{\mathrm{(L),I}},\varphi_{d_1^{\mathrm{II}}}^{\mathrm{(L),II}},\ldots,\varphi_{d_N^{\mathrm{II}}}^{\mathrm{(L),II}},\phi_0^{(\mathrm{L})}\right](z)}{\mathrm{W}\left[\varphi_{d_1^{\mathrm{I}}}^{\mathrm{(L),I}},\ldots,\varphi_{d_M^{\mathrm{I}}}^{\mathrm{(L),I}},\varphi_{d_1^{\mathrm{II}}}^{\mathrm{(L),II}},\ldots,\varphi_{d_N^{\mathrm{II}}}^{\mathrm{(L),II}}\right](z)} ~,
\label{eq:MILphi0} \\
&\phi_{\mathcal{D};0}^{(\mathrm{M,J})}(x) = [-2(1-y^2)]^{\frac{M+N}{2}}
\nonumber \\
&\quad \times\frac{\mathrm{W}\left[\varphi_{d_1^{\mathrm{I}}}^{\mathrm{(J),I}},\ldots,\varphi_{d_M^{\mathrm{I}}}^{\mathrm{(J),I}},\varphi_{d_1^{\mathrm{II}}}^{\mathrm{(J),II}},\ldots,\varphi_{d_N^{\mathrm{II}}}^{\mathrm{(J),II}},\phi_0^{(\mathrm{J})}\right](y)}{\mathrm{W}\left[\varphi_{d_1^{\mathrm{I}}}^{\mathrm{(J),I}},\ldots,\varphi_{d_M^{\mathrm{I}}}^{\mathrm{(J),I}},\varphi_{d_1^{\mathrm{II}}}^{\mathrm{(J),II}},\ldots,\varphi_{d_N^{\mathrm{II}}}^{\mathrm{(J),II}}\right](y)} ~.
\label{eq:MIJphi0}
\end{align}

The special cases 
[$\mathcal{D}^{\mathrm{I}}=\{ \ell \}$ and $\mathcal{D}^{\mathrm{II}}=\varnothing$] and 
[$\mathcal{D}^{\mathrm{I}}=\varnothing$ and $\mathcal{D}^{\mathrm{II}}=\{ \ell \}$] 
are called the type I and the type II $X_\ell$-Laguerre/Jacobi system, respectively. 
In Ref. \cite{Bougie:2018lvd}, the authors discussed the problem for the type II $X_1$-Laguerre system.

\subsection{The SWKB condition for the multi-indexed systems}
For the multi-indexed systems, the SWKB condition \eqref{eq:SWKB} reads
\begin{multline}
\int_a^b \sqrt{\mathcal{E}_{\mathcal{D};n}^{\mathrm{(M,\ast)}} - \left( \hbar\partial_x\ln\left|\phi_{\mathcal{D};0}^{\mathrm{(M,\ast)}}(x)\right| \right)^2}~\mathrm{d} x = n\pi\hbar \\
(n \in \mathbb{Z}_{\geqslant 0}) ~.
\end{multline}

For the case of the radial oscillator (L), with the groundstate eigenfunction \eqref{eq:MILphi0} and the energy eigenvalue  \eqref{eq:MIene} reduces to
\begin{equation}
I^{(\mathrm{M,L})} \coloneqq
\int_{a'}^{b'} \sqrt{4n - \left( \partial_{\xi}\ln\left|\phi_{\mathcal{D};0}^{(\mathrm{M,L})}(x)\right| \right)^2}~\mathrm{d}\xi = n\pi 
\label{eq:SWKB_ML}
\end{equation}
with $a'$ and $b'$ being the roots of the equation
\begin{equation}
\left( \partial_{\xi}\ln\left|\phi_{\mathcal{D};0}^{(\mathrm{M,L})}(x)\right| \right)^2 = 4n ~.
\end{equation}
Note that this formula \textit{does not depend upon} $\hbar,\omega$ but depends on $g$.

Similarly for the P\"{o}schl--Teller potential, the SWKB condition becomes, using the groundstate eigenfunction \eqref{eq:MIJphi0} 
and the energy eigenvalue \eqref{eq:MIene},
\begin{align}
&I^{(\mathrm{M,J})} \coloneqq \nonumber \\
&\int_{a'}^{b'} \sqrt{n(n+g+h) - (1-y^2)\left( \partial_y\ln\left|\phi_{\mathcal{D};0}^{(\mathrm{M,J})}(x)\right| \right)^2} \nonumber \\
&\hspace{4.5cm} \times\frac{\mathrm{d}y}{\sqrt{1-y^2}} = n\pi ~,
\label{eq:SWKB_MJ}
\end{align}
which \textit{is independent of} $\hbar$.
Our correct and explicit incorporation of the $\hbar$-dependency figure out that the integrals $I^{(\mathrm{M,\ast})}$ do not depend on $\hbar$ for both cases.

\subsection{Results}
\begin{figure}[t]
\centering
\includegraphics[scale=0.29]{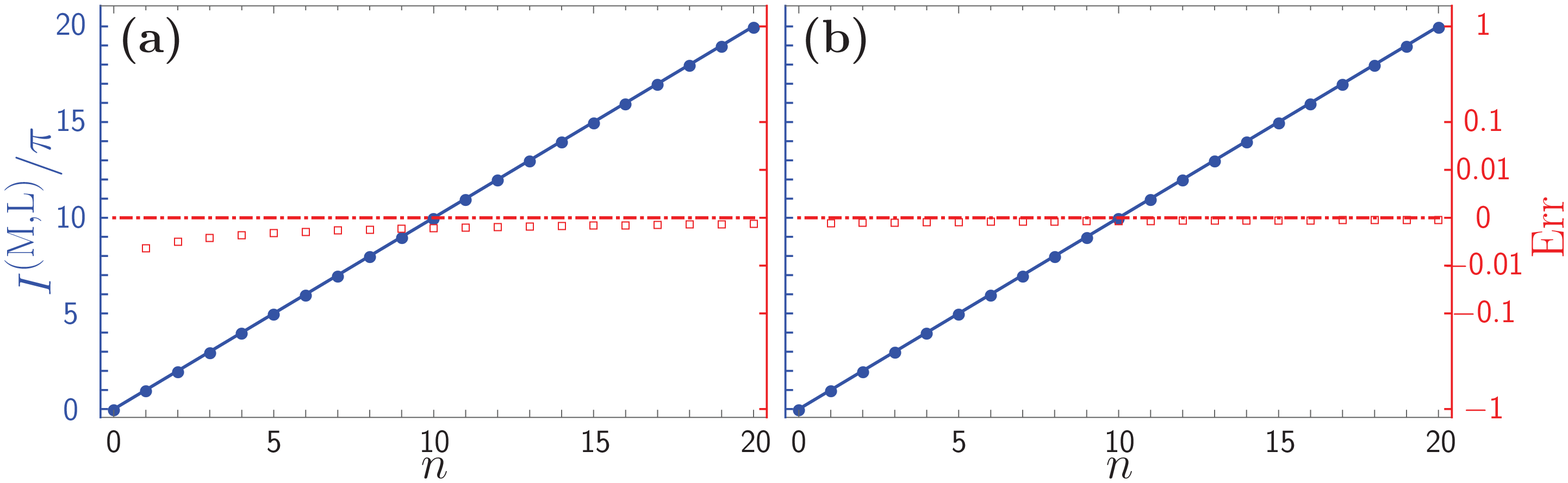}
\caption{The accuracy of the SWKB condition for the type II $X_1$-Laguerre system. The parameter $g$ is chosen 
as (a) $g=3$ and (b) $g=10$. 
The blue dots are the value of the integration $I^{(\mathrm{M,L})}$ (\ref{eq:SWKB_ML}) and the red squares are the corresponding  
errors defined by eq.~\eqref{eq:error}, while the blue line and the red chain line mean that the SWKB condition is exact and also $\mathrm{Err}=0~$.}
\label{fig:2ELX1}
\end{figure}

We calculate the SWKB conditions numerically to see the accuracy the SWKB conditions for the multi-indexed systems.
We plot the values of the integral $I^{(\mathrm{M},\ast)}/\pi$ and the relative errors for the conditions defined as
\begin{align}
\mathrm{Err} \coloneqq \frac{I^{(\mathrm{M},\ast)}-n\pi}{I^{(\mathrm{M},\ast)}}
\label{eq:error}
\end{align} 
where $n$ is the number of nodes.
In Fig.\ref{fig:2ELX1}, we show the result of the type II $X_1$-Laguerre system
\footnote{We adopt the rescaling for the 
plot of the error of as $\mathrm{sgn}(\mathrm{Err}) 2^{\log_{10}|\mathrm{Err}|}$.},  
which corresponds to the analysis of Ref. \cite{Bougie:2018lvd}.
The SWKB condition is not exact; while the errors are less than $10^{-2}$. 
For the larger parameter $g$, the error decreases and 
the condition becomes closer to the exact one. 
The claim of Ref. \cite{Bougie:2018lvd} still holds in our analysis, where the explicit $\hbar$-dependence is properly taken into account.

\begin{figure}[t]
\centering
\includegraphics[scale=0.29]{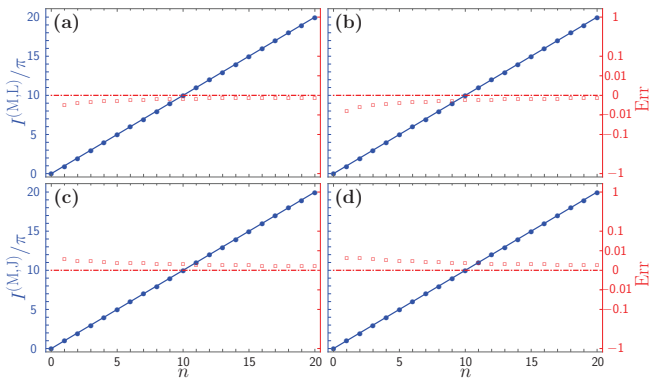}
\caption{The accuracy of the SWKB condition for the case of (a), (b) the multi-indexed Laguerre systems with $g=5$, 
and (c),(d) the multi-indexed Jacobi systems with $(g,h)=(5,6)$.
We choose $\mathcal{D}$ as 
(a),(c) 
[$\mathcal{D}^{\mathrm{I}}=\{ 1 \}$ and $\mathcal{D}^{\mathrm{II}}=\{ 2 \}$], 
and (b),(d)
[$\mathcal{D}^{\mathrm{I}}=\{ 1,2 \}$ and $\mathcal{D}^{\mathrm{II}}=\{ 2,3 \}$]. 
The blue dots are the value of the integration $I^{(\mathrm{M,L})}$ (\ref{eq:SWKB_ML}), $I^{(\mathrm{M,J})}$ (\ref{eq:SWKB_MJ}) and the red squares are the corresponding  
errors defined by eq.~\eqref{eq:error}, while the blue line and the red chain line mean that the SWKB condition is exact and also $\mathrm{Err}=0~$.
The maximal errors $|\mathrm{Err}|$ 
are (a) $9.7\times 10^{-4}$, (b) $4.7\times 10^{-3}$, (c) $1.3\times 10^{-3}$, (d) $2.2\times 10^{-3}$, respectively. }
\label{fig:MIL}
\end{figure}

Fig.\ref{fig:MIL} presents the typical examples of the analysis of the cases with the multi-indexed Laguerre and Jacobi polynomials. 
The plots are for the cases of 
[$\mathcal{D}^{\mathrm{I}}=\{ 1 \}$ and $\mathcal{D}^{\mathrm{II}}=\{ 2 \}$] and 
[$\mathcal{D}^{\mathrm{I}}=\{ 1,2 \}$ and $\mathcal{D}^{\mathrm{II}}=\{ 2,3 \}$], with appropriate choices
of parameters. The behaviors look similar in all cases; the maximal errors occur around the smaller $n$ with the values of orders $\sim 10^{-3}$. For larger $n$, it gradually reduces and in the limit $n\to\infty$, the SWKB condition will be restored. 
The Laguerre system is always underestimated, while the Jacobi is overestimated. 


\begin{figure}[h]
\centering
\includegraphics[scale=0.6]{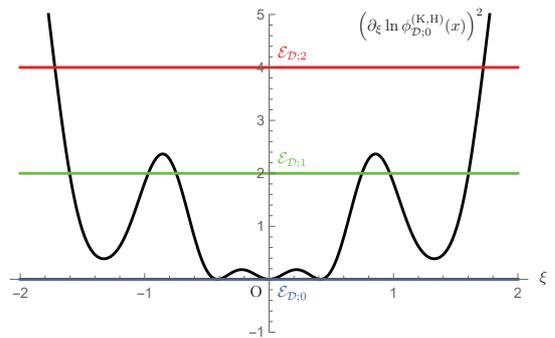}
\caption{The plot of the square of the superpotential $\left(\partial_{\xi}\ln\phi_{\mathcal{D};0}^{(\mathrm{K,H})}(x)\right)^2$ 
for $d=4$.
When $n=1$, this system has more than one set of turning points.}
\label{fig:W2}
\end{figure}


\section{Krein--Adler systems and the SWKB condition}
\label{sec:KA}

In this section, we examine the SWKB conditions for the systems with no shape invariance, 
so-called the Krein--Adler systems. 
From an exactly solvable Hamiltonian, one can construct infinitely many variants of exactly 
solvable Hamiltonians and their eigenfunctions by Krein--Adler transformations. 
The resulting systems are, however, not shape invariant, even if the starting systems are.
We study this class of systems in order to see the roles of the shape invariance, and the effects of SWKB in the exact solvabilities of the systems.


\begin{figure*}[t]
\centering
\includegraphics[scale=0.4]{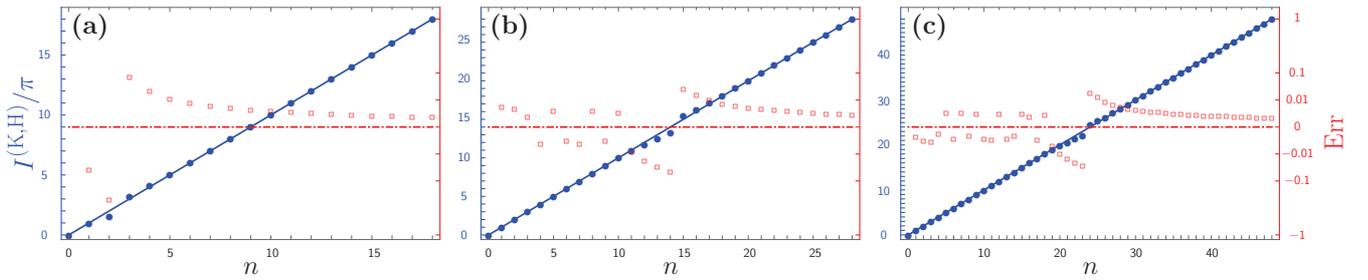}
\caption{The accuracy of the SWKB condition for the case of the Krein--Adler Hermite system~\eqref{eq:SWKB_KAH} 
for (a) $\mathcal{D}=\{ 3, 4 \}$, (b) $\mathcal{D}=\{ 15, 16 \}$ and (c) $\mathcal{D}=\{24, 25 \}$.
The blue dots are the value of the integration $I^{(\mathrm{K,H})}$ (\ref{eq:SWKB_KAH}) and the red squares are the corresponding  
errors defined by eq.~\eqref{eq:error} with $\mathrm{M}\to\mathrm{K}$, while the blue line and the red chain line mean that the SWKB condition is exact and also the $\mathrm{Err}=0~$.
The maximal errors $|\mathrm{Err}|$ are (a) $2.7\times 10^{-1}$, (b) $5.6\times 10^{-2}$, (c) $3.6\times 10^{-2}$, respectively.  }
\label{fig:KAH}
\end{figure*}


\subsection{Krein--Adler systems}

The Krein--Adler systems are obtained by the deformations of the three exactly solvable polynomials \eqref{eq:ES_ham}.
We choose the eigenfunctions with $n\in\mathcal{D}=\{d, d+1\}$ and $d\in\mathbb{Z}_{>0}$ as the seed solutions, deforming $\{ \phi_n(x) \}$ through the multiple Darboux transformation to obtain the Krein--Adler polynomials. 
Note that, during the transformation, the $d$th and $(d+1)$th eigenstates are deleted, and the new systems are no longer shape invariant. 
The resulting deformed systems read
\begin{equation}
\mathcal{H}_{\mathcal{D}}^{\mathrm{(K,\ast)}} \coloneqq \mathcal{H}^{(\ast)} - 2\hbar^2\partial_x^2\ln\left|\mathrm{W}\left[\phi_d^{(\ast)},\phi_{d+1}^{(\ast)}\right](x)\right|
\end{equation}
with
\begin{gather}
\mathcal{H}_{\mathcal{D}}^{\mathrm{(K,\ast)}}\phi_{\mathcal{D};\breve{n}}^{\mathrm{(K,\ast)}}(x) = \mathcal{E}_{\mathcal{D};\breve{n}}^{\mathrm{(K,\ast)}}\phi_{\mathcal{D};\breve{n}}^{\mathrm{(K,\ast)}}(x) ~, \\
\phi_{\mathcal{D};\breve{n}}^{\mathrm{(K,\ast)}}(x) = \frac{\mathrm{W}\left[\phi_d^{(\ast)},\phi_{d+1}^{(\ast)},\phi_{\breve{n}}^{(\ast)}\right](x)}{\mathrm{W}\left[\phi_d^{(\ast)},\phi_{d+1}^{(\ast)}\right](x)} ~,
\label{eq:KAene}
\end{gather}
where $\ast=\mathrm{H,L,J}$ and
\begin{equation}
\breve{n} \coloneqq \left\{
	\begin{array}{l}
	n \qquad (0 \leqslant n \leqslant d-1) \\
	n+2 \qquad (n \geqslant d)
	\end{array}
\right. \qquad (n \in \mathbb{Z}_{\geqslant 0})
\end{equation}
with the number of nodes $n$.
Especially for the groundstate ($n=0$),
\begin{align}
\phi_{\mathcal{D};0}^{(\mathrm{K,H})}(x) &= \mathrm{e}^{-\frac{\xi^2}{2}}\frac{\mathrm{W}\left[H_d,H_{d+1},1\right](\xi)}{\mathrm{W}\left[H_d,H_{d+1}\right](\xi)} ~,
\label{eq:KAHphi0}  \\
\phi_{\mathcal{D};0}^{(\mathrm{K,L})}(x) &= \mathrm{e}^{-\frac{z}{2}}z^{\frac{g+2}{2}}\frac{\mathrm{W}
\left[L_d^{(g-\frac{1}{2})},L_{d+1}^{(g-\frac{1}{2})},1\right](z)}{\mathrm{W}\left[L_d^{(g-\frac{1}{2})},L_{d+1}^{(g-\frac{1}{2})}\right](z)} ~,
\label{eq:KALphi0} \\
\phi_{\mathcal{D};0}^{(\mathrm{K,J})}(x) &= (1-y)^{\frac{g+2}{2}}(1+y)^{\frac{h+2}{2}} 
\nonumber \\
&\hspace{0.75cm}\times\frac{\mathrm{W}\left[P_d^{(g-\frac{1}{2},h-\frac{1}{2})},
P_{d+1}^{(g-\frac{1}{2},h-\frac{1}{2})},1\right](y)}{\mathrm{W}\left[P_d^{(g-\frac{1}{2},h-\frac{1}{2})},P_{d+1}^{(g-\frac{1}{2},h-\frac{1}{2})}\right](y)} ~,
\label{eq:KAJphi0}
\end{align}
respectively.


\begin{figure*}[t]
\centering
\includegraphics[scale=0.4]{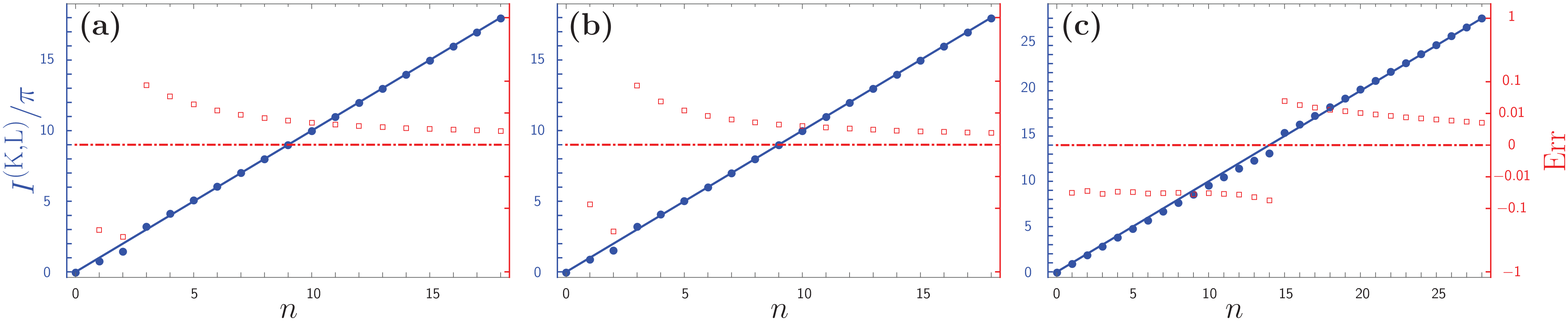}
\caption{The accuracy of the SWKB condition for the case of the Krein--Adler Laguerre system~\eqref{eq:SWKB_KAL} 
for (a) $g=3$, $\mathcal{D}=\{ 3, 4 \}$, (b) $g=30$, $\mathcal{D}=\{ 3, 4 \}$ and (c) $g=3$,  $\mathcal{D}=\{15, 16 \}$.
The blue dots are the value of the integration $I^{(\mathrm{K,L})}$ (\ref{eq:SWKB_KAL}) and the red squares are the corresponding  
errors defined by eq.~\eqref{eq:error} with $\mathrm{M}\to\mathrm{K}$, while the blue line and the red chain line mean that the SWKB condition is exact and also $\mathrm{Err}=0~$.
The maximal errors $|\mathrm{Err}|$ are (a) $3.4\times 10^{-1}$, (b) $2.8\times 10^{-1}$, (c) $6.5\times 10^{-2}$, respectively. }
\label{fig:KAL}
\end{figure*}



\begin{figure*}[t]
\centering
\includegraphics[scale=0.4]{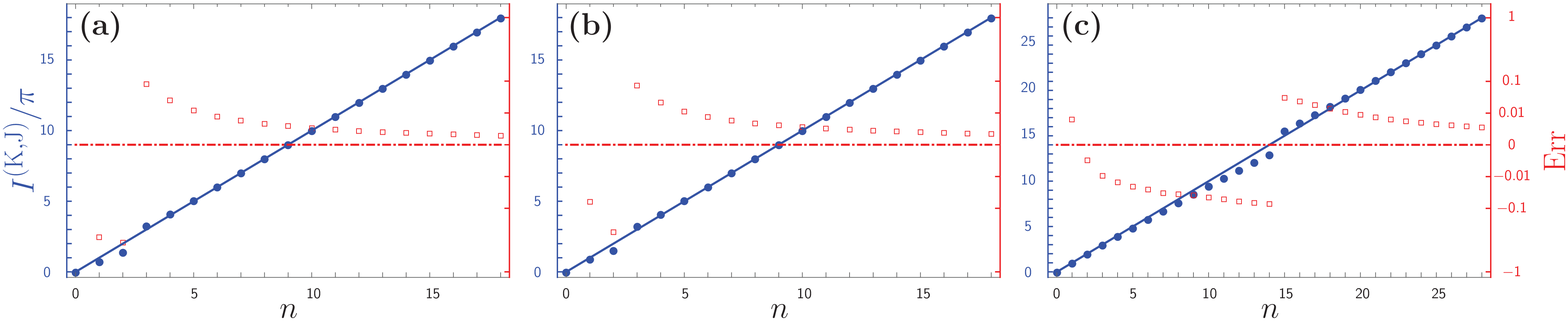}
\caption{The accuracy of the SWKB condition for the case of the Krein--Adler Jacobi system~\eqref{eq:SWKB_KAJ} 
for (a) $(g,h)=(3,4)$, $\mathcal{D}=\{ 3, 4 \}$, (b) $(g,h)=(30,40)$, $\mathcal{D}=\{ 3, 4 \}$ and (c) $(g,h)=(3,4)$,  $\mathcal{D}=\{15, 16 \}$.
The blue dots are the value of the integration $I^{(\mathrm{K,J})}$ (\ref{eq:SWKB_KAJ}) and the red squares are the corresponding  
errors defined by eq.~\eqref{eq:error} with $\mathrm{M}\to\mathrm{K}$, while the blue line and the red chain line mean that the SWKB condition is exact and also $\mathrm{Err}=0~$.
The maximal errors $|\mathrm{Err}|$ are (a) $4.2\times 10^{-1}$, (b) $2.9\times 10^{-1}$, (c) $8.1\times 10^{-2}$, respectively. }
\label{fig:KAJ}
\end{figure*}


\subsection{The SWKB conditions for the Krein--Adler systems}

For the Krein--Adler systems, the SWKB condition \eqref{eq:SWKB} reads
\begin{multline}
\int_a^b \sqrt{\mathcal{E}_{\mathcal{D};\breve{n}}^{\mathrm{(K,\ast)}} - \left( \hbar\partial_x\ln\left|\phi_{\mathcal{D};0}^{\mathrm{(K,\ast)}}(x)\right| \right)^2}~\mathrm{d} x = n\pi\hbar \\
(n \in \mathbb{Z}_{\geqslant 0}) ~.
\label{eq:SWKB_KA}
\end{multline}

For the case of the harmonic oscillator (H), with the groundstate eigenfunction \eqref{eq:KAHphi0} and the energy eigenvalue \eqref{eq:KAene}, 
Eq.~\eqref{eq:SWKB} reduces to
\begin{equation}
\int_{a'}^{b'} \sqrt{2\breve{n} - \left( \partial_{\xi}\ln\left|\phi_{\mathcal{D};0}^{(\mathrm{K,H})}(x)\right| \right)^2}~\mathrm{d}\xi = n\pi 
\label{eq:SWKB_KAHm}
\end{equation}
with $a'$ and $b'$ being the roots of the equation
\begin{equation}
\left( \partial_{\xi}\ln\left|\phi_{\mathcal{D};0}^{(\mathrm{K,H})}(x)\right| \right)^2 = 2\breve{n} ~.
\end{equation}
Unlike the conventional shape-invariant systems, however, this equation may possess more than two roots, i.e., there are sets of turning points: $\{ a'_i,b'_i \}$ with $a'_i\leqslant b'_i$~(see Fig.~\ref{fig:W2}).
The following prescription must be employed for the calculation of the integral in  (\ref{eq:SWKB_KAHm}). 
That is, the SWKB condition for the system is defined as the sum of the integrals in the l.h.s. of eq.~\eqref{eq:SWKB_KAHm} for all sets of turning points $\{ a'_i,b'_i \}$;
\begin{equation}
I^{(\mathrm{K,H})} \coloneqq
\sum_i \int_{a'_i}^{b'_i} \sqrt{2\breve{n} - \left( \partial_{\xi}\ln\left|\phi_{\mathcal{D};0}^{(\mathrm{K,H})}(x)\right| \right)^2}~\mathrm{d}\xi
=n\pi~.
\label{eq:SWKB_KAH}
\end{equation}
We note that $I^{\mathrm{(K,H)}}$ \textit{does not  depend upon} $\hbar, \omega$. 

For the radial oscillator (L) and the P\"{o}schl--Teller potential (J), the formulations are done in the same manner. 
With the groundstate eigenfunction of the radial oscillator \eqref{eq:KALphi0} and the energy eigenvalue \eqref{eq:KAene}, the SWKB condition is
\begin{equation}
I^{(\mathrm{K,L})} \coloneqq
\sum_i \int_{a'_i}^{b'_i} \sqrt{\breve{n} - z\left( \partial_z\ln\left|\phi_{\mathcal{D};0}^{(\mathrm{K,L})}(x)\right| \right)^2}~\frac{\mathrm{d}z}{\sqrt{z}}
=n\pi~.
\label{eq:SWKB_KAL}
\end{equation}
For the P\"{o}schl--Teller potential, using the groundstate eigenfunction~\eqref{eq:KAJphi0} and the energy 
eigenvalue \eqref{eq:KAene}, we obtain the SWKB condition
\begin{align}
&I^{(\mathrm{K,J})} \coloneqq
\nonumber \\
&\sum_i \int_{a'_i}^{b'_i} \sqrt{\breve{n}(\breve{n}+g+h) - \left( 1-y^2 \right)\left( \partial_y\ln\left|\phi_{\mathcal{D};0}^{(\mathrm{K,J})}(x)\right| \right)^2}
\nonumber \\
&\hspace{4.7cm}\times\frac{\mathrm{d}y}{\sqrt{1-y^2}}=n\pi~.
\label{eq:SWKB_KAJ}
\end{align}
$I^{(\mathrm{K,L})}$ and $I^{(\mathrm{K,J})}$ \textit{are also independent of} $\hbar$.
We can say the same thing as the case of the multi-indexed systems for the Krein--Adler systems: the integrals $I^{(\mathrm{K},\ast)}$ do not depend on $\hbar$ for all the cases after the $\hbar$-dependency is correctly and explicitly taken into account.

\subsection{Results}
We examine the SWKB conditions~\eqref{eq:SWKB_KAH}--\eqref{eq:SWKB_KAJ} numerically and show how accurate they are. 
We compute the cases of $\mathcal{D}=\{ 3, 4 \}$, $\{15, 16 \}$ and $\{24, 25 \}$ for the Hermite. We also calculate the cases of $\mathcal{D}=\{ 3, 4 \}$ and $\{15, 16 \}$ with the parameters $g=3,~30$ for the Laguerre, and with $(g,h)=(3,4),~(30,40)$ 
for the Jacobi polynomials. The results are shown in Fig.~\ref{fig:KAH} (Hermite), Fig.~\ref{fig:KAL} (Laguerre) and Fig.~\ref{fig:KAJ} (Jacobi). 
Notable feature of these results is that the maximum of the error occurs at the vicinity of the 
deleted levels and as moving away from the point, the error decreases and the SWKB condition tends to be exact. 
Also, the errors tend to be of opposite sign between the below and the above of the deleted levels. 
Note that the behavior at $n\to 0$ and $n\to \infty$ is not symmetrical, i.e.,  
for the smaller $n$, the value still decreases but seems not to go to the exact condition. 
For the larger value of the parameters $g,h$, 
it is expected that the error decreases, and we confirm the behavior (Figs.~\ref{fig:KAL}(b) and \ref{fig:KAJ}(b)). 
When we delete higher levels (larger $d$), 
we see different features. 
(See Figs.~\ref{fig:KAH}(b), \ref{fig:KAH}(c), \ref{fig:KAL}(c) and \ref{fig:KAJ}(c).)
Among these, the most distinctive behavior is seen in the Herimite polynomial. At below the deleted level, 
the integral value exhibits oscillating behavior around the exact one. 
The cases of the Laguerre, Jacobi polynomials are more moderate. 


\section{Discussion}
\label{sec:discussion}

As we have seen in Secs.~\ref{sec:MI} and \ref{sec:KA}, $\hbar$ is always factored out of the SWKB condition \eqref{eq:SWKB}, 
that is, the SWKB condition \textit{is totally independent of} $\hbar$, for the systems we have analyzed in this paper.
(See Eqs. \eqref{eq:SWKB_ML},\eqref{eq:SWKB_MJ},\eqref{eq:SWKB_KAH}--\eqref{eq:SWKB_KAJ}.)
Also, the SWKB condition cannot be derived from the WKB formalism because of the $\hbar$-dependency of superpotentials.
We have to say that the formal derivation of Eq.~\eqref{eq:SWKB} from Eq.~\eqref{eq:WKB} is just a fictitious, 
and the SWKB condition is a distinct condition from the WKB quantization condition.
Hence, the SWKB condition means nothing about semi-classical approximation.

The exactness of the SWKB condition is one independent criterion that can be considered in contrast to the WKB quantization condition;
the WKB quantization condition is exact for the case of the harmonic oscillator or the Morse potential, while the SWKB condition is exact for any conventional shape-invariant potentials.

The latter half of this section is devoted to the interpretation of our numerical results.
From the results of the multi-indexed polynomials (see Figs.~\ref{fig:2ELX1} and \ref{fig:MIL}), we found that the shape invariance was not the sufficient condition of the exactness of the SWKB condition. 
Thus, the shape invariance does not explain the exactness of the SWKB condition.

An interesting thing is that the SWKB conditions for all the systems we study are not exact, but near to the exact values. 
One may be curious about what guarantees the (approximate) satisfaction of the SWKB condition.
Figures~\ref{fig:KAH}--\ref{fig:KAJ} indicate that the whole distribution of the energy eigenvalues, or simply the level structure, 
guarantees the exactness of the SWKB condition.
The maximal errors are seen around the deleted levels $\mathcal{D}$ and the errors go closer to zero as $n \to \infty$ where the level structure of the systems are almost identical to that of the conventional shape-invariant ones.
The modifications of the conventional shape-invariant potentials change the level structures of the systems, and so does the values of the integral of the SWKB condition.
This is how the SWKB exactness breaks.

Since the SWKB condition is exact for all conventional shape-invariant potentials, this suggests that the SWKB integral characterize a system in terms of the net deviation from the conventional shape-invariant systems.
The breaks of the condition equation indicates the discrepancy between a system in question and the corresponding shape-invariant system.
Therefore, we propose that the analyses on other solvable systems through the SWKB condition is important to 
formalize the deviation from the conventional shape-invariant systems.

Before closing this section, we make a comment on the multi-indexed systems.
Since these systems are seen as the deletion of several ``negative'' levels from the conventional shape-invariant systems, a similar discussion to the above can be done.


\section{Conclusion}

In this paper, we studied the SWKB conditions for the new classes of exactly solvable systems with the multi-indexed Laguerre and Jacobi polynomials and the Krein--Adler Hermite, Laguerre and the Jacobi systems. 
The problem on their exactness of the SWKB conditions were already studied for 
the special case of the extended radial oscillator~\cite{Bougie:2018lvd} and the results indicated that the SWKB condition was not exact.  
However, in their analysis, the treatment of the $\hbar$ was peculiar and the results were problematical. 
We first studied the correct $\hbar$-dependency of the condition, and turned out that $\hbar$ can always be factored out of the SWKB condition.
Then, we numerically computed the integrals of the SWKB conditions, and the results clearly indicated that the conditions were always not exact but satisfied with some degree of accuracy for the systems we studied. 

This paper constitutes an initial step for a full understanding of the mathematical implications of the SWKB conditions. 
We now confident from our numerical studies that the SWKB condition is related to the deviation from conditional shape-invariant potentials, especially their level structures.
However, the mathematical explanation is still absent for this statement, and the following questions still remain:
\begin{itemize}

\item 
Several more solvable systems are known:
Investigation of the SWKB conditions for the conditionally exactly soluble potentials~\cite{Dutra} and the quasi-exactly-solvable potentials~\cite{Turbiner:2016aum}
with the parameter dependency may be valuable for the better understanding of the accuracy of the SWKB. 

\item 
The SWKB condition is only approximately satisfied in the new types of exactly solvable polynomials studied in this Letter. 
It would be a good challenge to identify the mechanism causing the discrepancy qualitatively.
We note here that the SWKB conditions for the conventional shape-invariant systems reduce to the proper quantization rules~\cite{serrano:2010}.
It is worth examining the proper quantization rule for the systems we investigated in this Letter.

\item
As mentioned in Sec.~\ref{sec:introduction},
our work is to be completed when the ``unknown parameter'' is identified.

\end{itemize}
These issues are currently studied and the results will be reported in our subsequent papers.

\appendix
\section{Erroneous analysis in Ref.~\cite{Bougie:2018lvd}}
In this Appendix, we discuss how the modeling of Ref.~\cite{Bougie:2018lvd} is not correct. 
The point is basic one; the system the authors considered is irrelevant to the known 
quantum mechanical problems such as the well-known radial oscillator for the explicit factor $\hbar$. 
They alleged that the additive shape invariance was realized for the parameters $a_i$ such that 
$a_{i+1}=a_i+\hbar$.
Their analysis was based on the expansion of the superpotential $W(a_i,\hbar)$ in power of $\hbar$, assuming 
that $W$ was independent of $\hbar$ except through the above shift of the parameter $a_i$. 
The main drawback in the analysis was that they overlooked the dependence of the parameter $a_i$ on $\hbar$.
Such wrong expansion with $\hbar$ inevitably leads to the devious result. 

Let us consider the eigenfunctions $\{\psi_n(x)\}$ of the system with the superpotential $W_0 = \dfrac{1}{2}\omega x - \dfrac{\ell}{x}$ in Ref.~\cite{Bougie:2018lvd}. 
The Schr\"{o}dinger equation is
\begin{align}
&\left[ -\hbar^2\frac{\mathrm{d}^2}{\mathrm{d}x^2} + {W_0}^2 - \hbar\frac{\mathrm{d}W_0}{\mathrm{d}x}  \right] \psi_n(x) \nonumber \\
= &\left[ -\hbar^2\frac{\mathrm{d}^2}{\mathrm{d}x^2} + \frac{\ell(\ell - \hbar)}{x^2} + \frac{1}{4}\omega^2x^2 - \frac{\omega}{2}(2\ell + \hbar) \right] \psi_n(x) \nonumber \\
= &2n\hbar\omega \psi_n(x) ~.
\end{align}
The eigenfunctions are obtained as
\begin{equation}
\psi_n(x) \propto
\mathrm{e}^{-\frac{\omega x^2}{4\hbar}}x^{\frac{\ell}{\hbar}}L_n^{\left(\frac{\ell}{\hbar}-\frac{1}{2}\right)}\left( \frac{\omega x^2}{2\hbar} \right) ~,
\label{eq:psi}
\end{equation}
where $L_n^{(\alpha)}(y)$ is the associated Laguerre polynomial.

On the other hand, many textbooks on quantum mechanics tell us the eigenfunctions of the radial oscillator $\{\phi_n(x)\}$
\begin{equation}
\phi_n(x) \propto
\mathrm{e}^{-\frac{\omega x^2}{4\hbar}}x^{\ell'+1}L_n^{\left( \ell' + \frac{1}{2} \right)}\left( \frac{\omega x^2}{2\hbar} \right) ~,
\label{eq:phi}
\end{equation}
with the potential $V(x) = \dfrac{\hbar^2\ell'(\ell'+1)}{x^2} + \dfrac{1}{4}\omega^2x^2$
and $\ell'$ is the angular momentum quantum number.
Here, Eq.~\eqref{eq:psi} and \eqref{eq:phi} \textit{must} coincide because both are the solutions of the same problem. 
As a result
\begin{equation}
\ell = \hbar(\ell' + 1) ~.
\end{equation}
which means that without $\ell$ being taken to be dependent upon $\hbar$, the model in Ref.~\cite{Bougie:2018lvd} \textit{never} 
be equivalent to the well-known radial oscillator. 
The modeling in Ref.~\cite{Bougie:2018lvd} is apparently not correct and the results are devious. 

\vskip 0.5cm\noindent
\begin{center}
{\bf Acknowledgment}
\end{center}
The authors would like to thank Ryu Sasaki for his careful and useful advice and comments.  
We also appreciate Naruhiko Aizawa, Yuki Amari, Atsushi Nakamula for valuable discussions. 
Y.N. thanks Yukawa Institute for Theoretical Physics, Kyoto University. Discussions during the YITP workshop YITP-W-19-10 on ``Strings and Fields 2019'' 
were useful to complete this work. N.S. was supported in part by JSPS KAKENHI Grant Number JP 16K01026 and B20K03278(1).
\bibliography{SWKB_bib.bib}

\end{document}